# High-brightness Graphite Photocathodes

Po-Hsun Wu,[1] Yu-Chieh Lo,[2] and Yen-Chieh Huang,[1,2,a)]

**AFFILIATIONS**

[1] Department of Physics, National Tsing Hua University, Hsinchu 30013, Taiwan.

[2] Institute of Photonics Technologies, National Tsing Hua University, Hsinchu 300044, Taiwan

[a)] Author to whom correspondence should be addressed: ychuang@ee.nthu.edu.tw

**ABSTRACT**

A robust and infrared laser-excited photocathode with high quantum efficiency, high brightness, and low cost, operating under a moderate vacuum, has long been sought by the accelerator and microscopy communities. This study investigates various types of graphite photocathodes, including bulk, sheet, and flake graphite, in the regime of thermionically assisted photoemission by an irradiating infrared laser. Our experiment reveals that, under space-charge-limited photoemission, the flake-graphite photocathode with a dense population of nano-graphene fins on its surface exhibits the highest quantum efficiency, which is 770 times greater than that of a copper photocathode irradiated by the same infrared laser at 1064 nm. With our theory considering both thermionic and multiphoton emissions, we determine that the flake graphite photocathode is 200 times brighter than a copper photocathode irradiated by an infrared laser and is as bright as a LaB$_6$ field emitter.

## I. INTRODUCTION

A photocathode is a surface-emission device that converts incident photons into emitted electrons. For instance, a photomultiplier or an image intensifier employs one or more photocathodes to receive photons and emit electrons with subsequent amplification. An ultrafast electron microscope with its cathode gated by short laser pulses provides both spatial and temporal resolutions for nano-imaging[1]. A high-energy photoinjector, irradiated by an infrared laser, is useful for studying electron radiation[2] and diffraction[3] with temporal resolution. High brightness, high quantum efficiency, low cost, small work function, and robustness are desirable for a photocathode. In particular, a low work function permits the use of a less complex and low-cost laser to enable photoemission. Copper is a popular photocathode material often installed in a MeV photoinjector, operating under a high vacuum with a quantum efficiency of $10^{-5}$~$10^{-6}$, when excited by an ultraviolet (UV) laser[2]. A laser is more efficient, stable, and robust in the near-infrared spectrum. The work function of copper[4] is in the range of 4.53-5.10 eV, which provides negligible quantum efficiency with infrared laser irradiation. A Cs$_2$Te photocathode[5] may have 10-20% quantum efficiency when excited by a visible-UV laser, although it requires delicate preparation and a high vacuum to operate. Semiconductor materials such as GaAs[6,7] and silicon nano-tip structures[8] also exhibit high quantum efficiencies when irradiated by near-infrared or visible lasers. However, semiconductor photocathodes are mostly susceptible to laser damage and require an ultrahigh vacuum for operation.

Graphite is a low-cost material that is commonly used for battery electrodes. Unlike copper, which melts slightly above 1000°C and has a reflective surface, graphite has a high sublimation point[9] of approximately 3600°C and a broad absorption band in the visible and infrared spectra. In the past, graphite has been used as an explosive cathode[10] for generating an intense electron current in a high-power microwave system. Graphite's chemical inertness, high sublimation temperature, ease of fabrication, wide availability, and moderate vacuum condition for electron emission make it a promising and robust candidate for a cathode material. To the best of our knowledge, graphite has never been studied for its use as a high-current photocathode. The large work function[11], although slightly lower than that of copper, is perhaps the reason for it not being considered a promising photoemitter with infrared laser irradiation. However, the work function is not the only factor that determines the quantum efficiency and brightness of photocathodes. Laser-induced heating and field emission on the cathode material are also crucial for the operation of photocathodes. In this work, we exploit different types of graphite, including bulk, sheet, and so-called flake graphite, as photocathodes by illuminating them with lasers having photon energy significantly lower than the work function of graphite. We compared the performance of graphite photocathodes with that of copper. Flake graphite is a new type of graphite fabricated using randomly distributed graphene nanoflakes on a metal surface.

The remainder of this paper is organized as follows. In Sec. II, we first introduce the photoemission theory developed by Fowler[12], DuBridge[13,14], and Bechtel[15,16]. To account for both thermionic and multiphoton emissions within a single cathode material, we also present a new theory for calculating the mean transverse energy of the emitted electrons. In Sec. III, we present the materials and methods for generating and measuring photocurrents from graphite and copper. In Sec. IV, we first studied the performance of green- and infrared laser-irradiated bulk-graphite photocathodes and then reported the performance of infrared laser-irradiated sheet- and flak-graphite photocathodes. We fit the experimental data with the theories to determine the electron emission mechanisms, quantum efficiency, and brightness of the graphite photocathodes in comparison with a copper photocathode. Finally, we conclude the paper in Sec. V.

## II. THEORY

Fowler developed a temperature-dependent photoemission model based on the Fermi-Dirac distribution for single-photon absorption by introducing the Fowler function to describe this phenomenon. DuBridge further validated Fowler's theory and investigated the energy distribution of the photoelectrons. Bechtel expanded the theories to include both thermionic emission (zero photon with $n = 0$) and multiphoton emission ($n$ photons with $n = 1, 2, 3...$). In this research, we fit the experimental data with the theories developed by Fowler, DuBridge, and Bechtel to determine the electron emission mechanisms of various graphite photocathodes. In the following section, we first summarize the Fowler-DuBridge-Bechtel theory. To determine the intrinsic emittance of the emitted electrons, we present a new theory for calculating the mean transverse energy of electrons emitted from a thermionically assisted photoemission process.

The mechanisms of the electronic transition from a bound state to a free state above the vacuum level include thermionic emission, photoemission, and field emission. The Fowler-DuBridge-Bechtel theory expresses the emission current of a photocathode as a summation of all the partial currents $J_n$ with $n$ being the number of photons involved in the emission process for generating $J_n$, given by

$$J = \sum_{n=0}^{\infty} J_n. \tag{1}$$

The specific expression of $J_n$ is written as

$$J_n = a_n A_0 \left(\frac{e}{h\nu}\right)^n (1-R)^n I^n T^2 F\left[\frac{nh\nu - (\Phi - \Delta\Phi_{Schottky})}{k_B T}\right], \tag{2}$$

where $a_n$ is an amplitude coefficient, $R$ is the reflectance of the cathode surface, $I$ is the incident laser intensity, $T$ is the temperature, $A_0 = \frac{4\pi e m_e k_B}{h^3}$ is the Richardson constant, $m_e$ is the electron mass, $e$ is the electron charge, $k_B$ is the Boltzmann constant, $h$ is the Planck constant, $h\nu$ is the photon energy, $F[\mu]$ is the so-called Fowler function with $\mu = \frac{nh\nu - \Phi + \Delta\Phi_{Schottky}}{k_B T}$, $\Phi$ is the work function of the cathode material, and $\Delta\Phi_{Schottky}$ is the shift in the work function due to the Schottky effect. The effective work function of a cathode under a biased electric field is therefore $\Phi - \Delta\Phi_{Schottky}$. For our case, $R = 0$, because graphite strongly absorbs a laser at 532 nm or 1064 nm.

In our experiment, to irradiate the photocathodes, we adopted a Q-switched Nd:YAG laser producing 20-ns pulses at 1064 nm and 14-ns pulses at 532 nm. According to the so-called two-temperature model[17], thermal equilibrium is reached between electrons and phonons in graphite due to the long duration of our laser pulses. Therefore, the Debye model[18,19] allows us to approximate the specific heat of graphite as a constant when the temperature exceeds the Debye temperature of graphite[20,21] at 413 K. This approximation simplifies Eq. (2) for our data fitting, because the temperature change of the sample becomes linearly proportional to the laser intensity. We can now write the temperature as $T = \Delta T + T_0 = \eta I + T_0$, where $\Delta T = \eta I$ with $\eta$ being the temperature conversion constant and $T_0$ is the room temperature. For a cathode involving thermionic emission, $T \gg T_0$, the laser intensity is simply $I \sim T/\eta$. The equation for the photocurrent density associated with a $n$-photon emission process is reformulated as

$$J_n = a_n A_0 \eta^{-n} T^{2+n} F\left[\frac{nh\nu - (\Phi - \Delta\Phi_{Schottky})}{k_B T}\right] \tag{3}$$

with $a_n$, $\eta$, and $T$ being the fitting parameters.

To infer the brightness of an electron source, one needs to calculate the intrinsic emittance of the electron current. A high-quality electron beam has a small emittance value. The geometric emittance of an electron beam ($\epsilon_{RMS}$) is defined as $\epsilon_{RMS} = \sqrt{\langle x^2 \rangle \langle x'^2 \rangle - \langle x \cdot x' \rangle^2}$, where $x$ and $x'$ are the transverse position and the emitting angle of an electron, respectively, and $\langle A \rangle$ denotes averaging for the quantity $A$ over many electrons. With negligible space-charge fields in the linear-emission regime, the angle-position correlation $\langle x \cdot x' \rangle$ can be ignored for electrons emitted from a planar cathode. The expression of the geometric emittance reduces to $\epsilon_{RMS} \sim \sigma_x \sigma_{x'}$, where $\sigma_x$ and $\sigma_{x'}$ are the standard deviations of the electron position and angle, respectively. For a high-energy accelerator, the normalized emittance is defined as $\epsilon_N = \beta_z \gamma \epsilon_{RMS}$, where $\beta_z$ is the electron longitudinal speed normalized to the vacuum speed of light $c$ and $\gamma$ is the Lorentz factor. Normalized emittance is an invariant quantity under particle acceleration. For our case with $\gamma \sim 1$, the normalized emittance is given by:

$$\epsilon_N \sim \sigma_x \frac{\sqrt{\langle v_x^2 \rangle}}{c} = \sigma_x \sqrt{\frac{MTE}{m_e c^2}}, \tag{4}$$

where $MTE$ is the mean transverse energy of the electrons.

For pure thermionic emission, the $MTE$ is known to be the average thermal energy of a particle in an ideal gas or $\sim k_B T$[22,23]. For pure multiphoton emission of electrons from a cold cathode, the *MTE* approaches the value[24],

$$MTE \to \frac{nh\nu - \Phi}{3}, \quad (5)$$

which is independent of temperature. However, for our case, the photoemission from graphite is assisted by thermionic emission, because the strongly absorbed laser energy raises the surface temperature of the graphite cathode. Based on Fowler's theory and an assumed Fermi-Dirac distribution of an electron gas for a cathode material, we derive a theory for calculating the *MTE* of the electrons near the emission threshold for thermionically assisted photoemission[25]:

$$MTE = k_B T \frac{W[\mu]}{F[\mu]}, \quad (6)$$

where

$$W[\mu] = \begin{cases} -\sum_{k=1}^{\infty} (-1)^{k-1} \frac{e^{k\mu}}{k^3} & , \mu \leq 0 \\ \frac{1}{6}\mu(\mu^2 + \pi^2) - \sum_{k=1}^{\infty} (-1)^{k-1} \frac{e^{-k\mu}}{k^3} & , \mu \geq 0 \end{cases} \quad (7)$$

In the high-temperature limit, $T \to \infty$, thermionic emission dominates the electron-generation mechanism of a cathode. As expected, Eq. (6) converges to the average thermal energy of a thermal partile $MTE \to \frac{9\zeta(3)}{\pi^2} k_B T \approx 1.1 k_B T$, where $\zeta(s) = \sum_{k=1}^{\infty} k^{-s}$ is the Riemann zeta function. For finite *T*, thermionic emission could still dominate, if the *n*-photon energy is too small to overcome the work function of a cathode material ($\mu \leq 0$). For such a case, the first expression of Eq. (7) properly gives the convergence, $MTE \to k_B T$. On the other hand, in the limit of pure photoemission, $\mu > 0$ and $T \to 0$, Eq. (6) converges to the well-known *MTE* value in Eq. (5) for cold-cathode photoemission.

The brightness of an electron beam describes the usefulness of an electron source for various applications, given by

$$\boldsymbol{B} = \frac{I}{\varepsilon_x \varepsilon_y}, \quad (8)$$

where $I$ is the beam current, and $\boldsymbol{\varepsilon_x}$ and $\boldsymbol{\varepsilon_y}$ are the emittance in the *x* and *y* directions, respectively. For the circular beam in our experiment, $\boldsymbol{\varepsilon_x} = \boldsymbol{\varepsilon_y}$ due to symmetry. In Sec. 4, we will use Eqs. (6, 7) with theoretically fitted experimental data to calculate the electron emittance and use Eq. (8) to infer the brightness of our graphite photoemitters.

### III. MATERIALS AND METHODS

Three different graphite samples were prepared for this study. The first sample, referred to as bulk graphite, is a cylindrical rod made of pure graphite, with a diameter of 3 mm and a length of 10 mm, secured by a drill chuck in a copper housing. The second sample, referred to as sheet graphite, is a high-purity planar graphite sample, measuring an area of 15 mm × 12 mm and a thickness of 46 μm, securely attached to a copper electrode. The third sample, flake graphite, is formed by applying a thin coating of graphite spray (KONTAKT KT-33L) onto a copper substrate of 15 mm × 12 mm, followed by drying at 500°C for 1 minute, resulting in a 10 μm thick layer. The KONTAKT KT-33L spray was originally designed for lubri¬cating mechanical parts. The scanning-electron-microscope (SEM) images of the samples in Figs. 1(a-c) reveal the morphology of bulk graphite, sheet graphite, and flake graphite, respectively. The bulk graphite in (a) is typically fabricated from fine grinding and hot pressing of graphite powders, showing little crystalline structure. The sheet graphite in (b) contains large and stacked graphene domains on a planar surface. The flake graphite in (c) contains many sharp edges and nano-fins of broken graphene domains, distributed over a 3-dimensional surface.

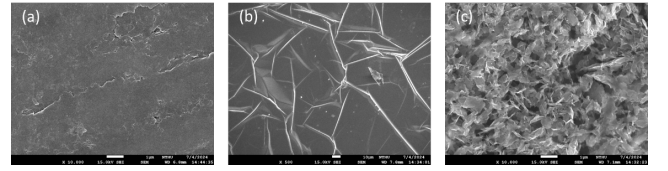

**FIG. 1.** SEM images of (a) bulk graphite with little crystalline structure, (b) sheet graphite with stacked graphene domains, and (c) flake graphite with 3-dimensionally distributed graphene flakes

A single-crystal graphite is formed from stacks of graphene layers loosely bonded by the Van der Waals force. The in-plane mobility of electrons in a stack of graphene layers is higher than the out-of-plane mobility of electrons. We expect that the work function of a graphene layer is anisotropic. The surfaces of our graphite samples contain different sizes of graphene domains oriented in different directions. By using ultraviolet photoelectron spectroscopy (UPS), we deduced 3.77, 3.85, and 3.66 eV for the work functions of the bulk, sheet, and flake graphite, respectively. As expected, flake graphite, comprising many protruding nano-fins of broken graphene on its surface, has the lowest work function compared with that of bulk and sheet graphite. Sheet graphite has the largest work function due to the lower out-of-plane electron mobility for photoemission. All the measured values are approximately ~1 eV lower than the known value for copper, permitting the use of a near-UV laser, such as the 3rd harmonic of a popular Nd or Yb laser, for inducing one-photon emission from a graphite photocathode. However, this study is to exploit the potential of using a much simpler infrared or visible laser to drive a graphite photocathode.

Fig. 2 depicts the schematic of our experimental setup for measuring the photocurrent from different graphite samples. The graphite cathode was electrically attached to a planar electrode, with the emission surface kept at a constant distance from the anode for all the measurements. The experimental chamber had a vacuum pressure of 10[-6] torr, which is easily achievable for an ordinary vacuum system. The cathode was biased at a DC voltage of -40 kV, while the anode was set to ground. The excitation laser is a Q-switched Nd:YAG laser, having a wavelength of 1064 nm and a pulse width of 20 ns. To study visible laser-induced photoemission of graphite, a second-harmonic generator was used to convert the infrared laser to a green laser at 532 nm. The green-laser pulse width is 14 ns. We injected the laser pulse through a transmission hole in a homemade Faraday cup and then another transmission hole in the anode to reach the cathode at normal incidence. Upon laser irradiation, electrons are emitted from the cathode and accelerated toward the anode. Given a 2-cm distance between the cathode and anode, the acceleration gradient for the photoelectrons is 2 MV/m. As soon as the electrons pass through the same laser apertures in the anode, a permanent magnet bends the electrons toward the conducting wall of the transmission hole in the Faraday cup. Currents flowing into both the anode and the Faraday cup are monitored by an oscilloscope, permitting alignment and analysis of the electron current.

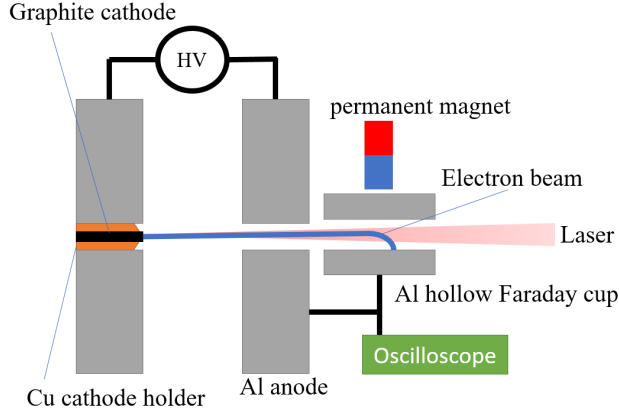

**FIG. 2.** Schematic of the experimental setup for characterizing different photocathode samples. The laser pulse passes through a hole in a Faraday cup and then another hole in an anode to reach the surface of a photocathode sample in a copper electrode. The accelerated electrons are bent by a permanent magnet and collected by the side wall of the Faraday cup for current measurements.

## IV. RESULTS

In the following, we first study the bulk-graphite photocathode irradiated by the pulse laser at 1064 and 532 nm. We then present the performance of different graphite photocathodes irradiated by the 1064-nm laser, in comparison with that of a copper photocathode.

### A. Bulk-graphite Photocathode

Fig. 3 shows the measured photocurrent density of bulk graphite excited by our 1064-nm laser and the corresponding fitting curves for $J_n$ with $n$ varying from 0 for pure thermionic emission to 4 for thermionically assisted 4-photon emission. Table 1 enumerates the fitting parameters for $J_n$. In Fig. 3, only $J_0$, $J_1$, and $J_2$ terms provide satisfactory fitting to the experimental data. However, the maximum temperatures of the graphite surface under laser irradiation for $J_0$ and $J_1$ have already exceeded the sublimation point of graphite at about ~3900 K. The $J_2$ term emerges as the most probable term responsible for the measured electron emission, indicating a thermionically assisted 2-photon emission with a maximum cathode temperature of about 2400 K near saturation. When the irradiated laser intensity reaches ~7.5 MW/cm², the photocurrent saturates at ~6.5 A/cm². Over the range of the laser intensity in Fig. 3, we observed no laser-induced damage on the cathode surface. We attribute the saturation of the photocurrent to the space-charge effects. From this point forward, we present only the measured photocurrent to the onset of saturation.

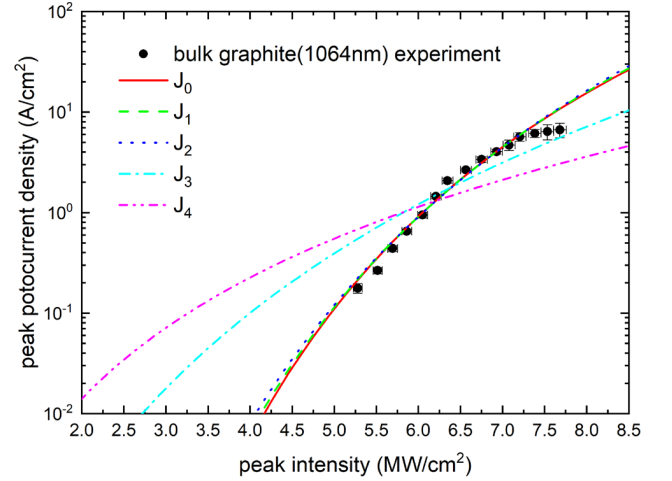

**FIG. 3.** Measured photocurrent density under different laser intensities and the theoretical fitting curves for bulk graphite under 1064-nm laser irradiation. The $J_0$, $J_1$, and $J_2$ terms fit well to the experimental data, but only the $J_2$ term has a maximum temperature below the sublimation point of graphite.

**TABLE I.** Fitting parameters for the measured photocurrent in Fig. 3. The $J_2$ term is the most probable term to explain the electron emission from the bulk-graphite cathode irradiated by the 1064-nm laser.

|  | $a_n [(\frac{cm^2}{MW})^n]$ | $\eta [\frac{Kcm^2}{MW}]$ | Max. T [K] | Fitting Variance |
|---|---|---|---|---|
| $J_0$ | 2.80×10² | 693 | 5627 | 0.123 |
| $J_1$ | 1.79×10¹ | 513 | 4241 | 0.133 |
| $J_2$ | 1.57 | 273. | 2400 | 0.142 |
| $J_3$ | 1.26×10⁻³ | 76.2 | 885 | 2.15 |
| $J_4$ | 1.26×10⁻¹¹ | 1.03 | 308 | 4.73 |

Regarding the electron emission from the bulk graphite excited by the 532-nm laser, Fig. 4 presents the measured photocurrent density fitted with curves of $J_n$ with $n$ varying from 0 to 2. In the figure, only the $J_0$ and $J_1$ terms provide satisfactory fitting to the experimental data. Given the fitting parameters in Table 2, it is seen that the maximum temperature under the laser irradiation for $J_0$ has already surpassed the sublimation point of graphite. Therefore, $J_1$ is the most probable term for the 532-nm-laser excited electron emission from the bulk graphite. In other words, the emission mechanism of the laser-irradiated bulk graphite at 532 nm is thermionically assisted 1-photon emission with a maximum cathode temperature of about 3000K near saturation. We purposely plot Figs. 3 and 4 on the same scale for comparison. Under 532-nm laser irradiation, the graphite cathode begins emitting photoelectrons at a much lower laser intensity of 3.1 MW/cm². The photocurrent saturates at a value of 2.5 A/cm², when the laser irradiating intensity is just 4.5 MW/cm².

Under microscope inspection, we observed the surface damage to the graphite surface. Therefore, the saturation mechanism of the photoemission from this green-laser-excited graphite cathode is laser-heating-induced material damage.

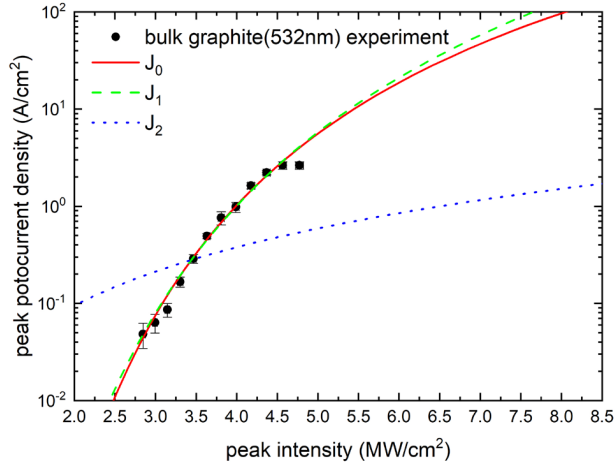

**FIG. 4.** Measured photocurrent density under different laser intensities and the theoretical fitting curves for bulk graphite under 532-nm laser irradiation. The $J_0, J_1$ terms fit well to the experimental data, but the $J_0$ term has a maximum temperature exceeding the sublimation point of graphite.

**TABLE II.** Fitting parameters for the measured photocurrent in Fig. 3. The $J_2$ term is the most probable term to explain the electron emission from the bulk-graphite cathode irradiated by the 1064-nm laser.

| | $a_n[(\frac{cm^2}{MW})^n]$ | $\eta[\frac{Kcm^2}{MW}]$ | Max. T [K] | Fitting Variance |
|---|---|---|---|---|
| $J_0$ | 68.4 | 1562 | 7751 | 5.16×10$^{-2}$ |
| $J_1$ | 9.86 | 572 | 3027 | 6.05×10$^{-2}$ |
| $J_2$ | 4.96×10$^{-10}$ | 0.886 | 304 | 3.45 |

Quantum efficiency ($QE$) is a figure of merit for a photocathode, describing the probability of converting an incident photon into an emitted electron from the cathode, given by

$$QE = \frac{(number\ of\ photoelectrons)}{(number\ of\ photons)} = \frac{J/e}{I/h\nu}. \quad (9)$$

For thermionically assisted photoemission, quantum efficiency depends on the cathode temperature and therefore on the irradiating laser intensity. Fig. 5 shows (a) the cathode temperature and (b) quantum efficiency of the bulk graphite cathode versus the irradiated laser intensities, plotted by using the parameters of the fitted $J_2$ and $J_1$ terms for the 1064- and 532-nm-laser-induced photoemission, respectively.

The high temperature of the green-laser irradiated bulk graphite is due to the strong absorption of graphite in the visible spectrum[26]. The quantum efficiency of the bulk graphite at 1064 nm saturates at $1 \times 10^{-6}$ with $I = 7.5$ MW/cm$^2$ due to the space-charge effects, whereas that at 532-nm saturates at $6.4 \times 10^{-7}$ with $I = 4.5$ MW/cm$^2$ due to laser-heating-induced material breakdown. The temperature conversion efficiency $\eta$ at 1064 nm, 273.4 [$\frac{Kcm^2}{MW}$], is lower than that at 532nm, 571.7 [$\frac{Kcm^2}{MW}$]. The fast heating from the 532-nm laser enables the one-photon photoemission from bulk graphite. The two-photon emission of bulk graphite induced by the 1064-nm laser has a higher threshold irradiation intensity due to a lower temperature at the cathode surface. The $10^{-6}$ quantum efficiency of the bulk-graphite photocathode is comparable to that of a copper photocathode irradiated by a UV laser at ~250 nm[27] .

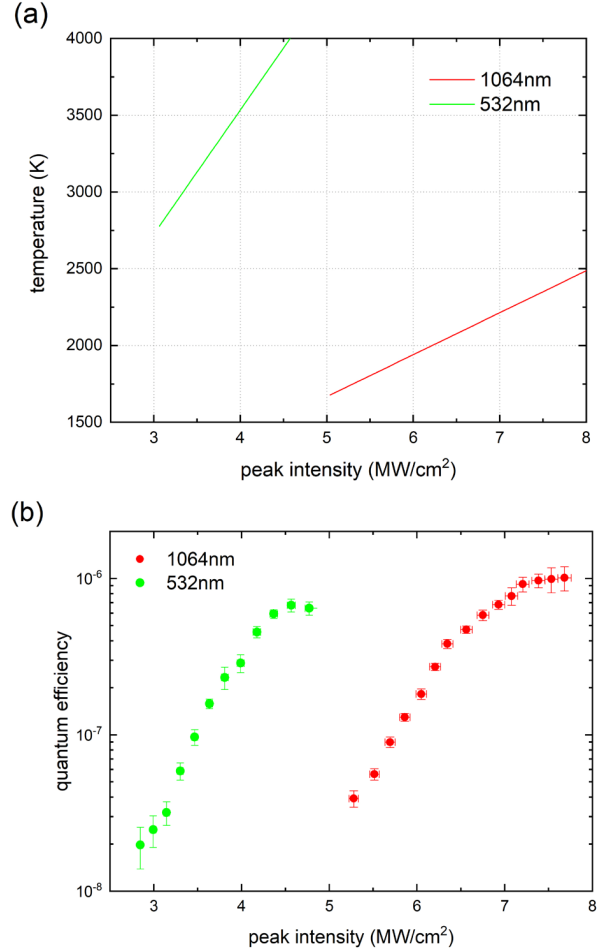

**FIG. 5.** (a) Cathode temperature and (b) quantum efficiency (QE) vs. laser intensity for bulk graphite irradiated by the 1064-nm laser (red) and the 532 nm laser (green). At saturation, the QE is approximately 10$^{-6}$, which is comparable to that of a copper photocathode irradiated by a UV laser at ~250 nm.

The emittance of an electron beam depends on MTE, which is a function of cathode temperature and, consequently, laser intensity. Fig. 6 plots (a) the spot-normalized emittance, $\epsilon_N/\sigma_x$ , calculated from Eq. (4) and (b) beam brightness calculated from Eq. (8) versus laser intensity for the 1064-nm-laser and 532-nm-laser irradiated bulk-graphite photocathode.

Owing to fast heating, the thermal emittance of the graphite photoemitter excited by the 532-nm laser is relatively higher. Furthermore, the 532-nm laser causes cathode damage at a low laser intensity. As a result, the maximum brightness of the infrared-laser-irradiated bulk-graphite photoemitter was approximately 20 times higher than that of the green-laser-irradiated emitter, even though the maximum photocurrent of the infrared-laser-irradiated emitter was only 2-3 times higher. The space-charge-limited brightness of the 1064-nm-laser-irradiated graphite emitter reaches 2.8 A/(mm·mrad)$^2$ at an irradiating laser intensity of 7.7 MW/cm$^2$.

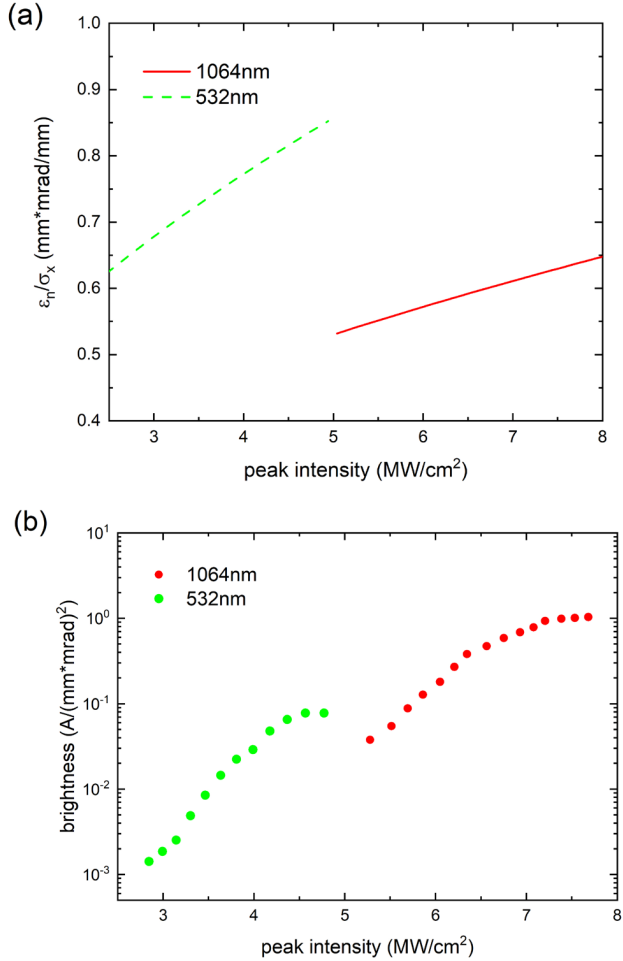

FIG. 6. (a) Spot-normalized emittance and (b) brightness of the bulk graphite photocathode vs. incident laser intensity at 1064 nm (red) and 532 nm (green). Owing to the high thermal emittance and low material-damage threshold, at saturation, the maximum brightness of the green-laser irradiated bulk-graphite emitter is about 20 times smaller than that of the infrared-laser irradiated one.

In summary, the strong absorption of graphite at 532 nm has a few consequences. Compared with a graphite photocathode irradiated by an infrared laser, a graphite photocathode irradiated by a green laser of the same intensity has a lower emission threshold, higher surface temperature, higher thermal emittance, and lower maximum brightness. Furthermore, a 532-nm laser tends to cause material damage at a low laser intensity. For what follows, we focus our study on using a 1064-nm laser to irradiate different graphite photocathodes and compare their performance with that of a copper photocathode

### B. Sheet- and Flake-graphite Photocathodes

Using the same experimental setup shown in Fig. 2, we continued measuring the photoemission from sheet- and flake-graphite photocathodes irradiated by the Q-switched Nd:YAG laser at 1064 nm. Fig. 7 presents the measured and fitted photocurrent density versus incident laser intensity at 1064 nm for all the graphite photocathodes. On the same plot, we also show the data from a copper photocathode for comparison. The data points are only presented up to the onset of current saturation. Interestingly, flake graphite exhibits the lowest threshold for electron emission and the highest saturation photocurrent density. In contrast, sheet graphite has a relatively high threshold for electron emission due to its lower out-of-plane electron mobility. Under the same experimental setup, the photocurrent from a copper cathode is ~2 orders of magnitude lower. To understand the emission mechanism, we again perform curve fitting by adopting $J_n$ with $n$ varying from 0 to 4. For sheet graphite and flake graphite, only $J_0$, $J_1$, and $J_2$ provide satisfactory fitting for the measured photocurrents.

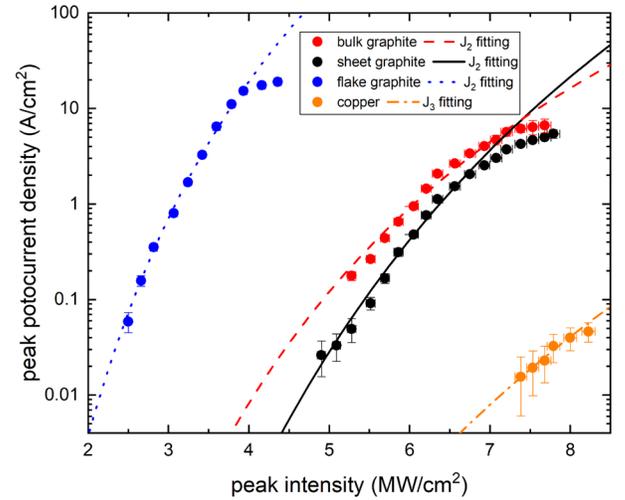

FIG. 7. Measured data and the best fitted curves for the photocurrent density vs. laser intensity at 1064 nm for bulk-graphite, sheet-graphite, flake-graphite, and copper photocathodes. Flake graphite exhibits the lowest emission threshold and highest saturation current. The maximum photocurrent from copper is approximately two orders of magnitude lower than that from graphite.

However, the cathode temperatures under maximum laser irradiation for $J_0$ and $J_1$ exceed the sublimation temperature of graphite. Similar to bulk graphite, $J_2$ emerges as the most probable term to explain the emission mechanism of the sheet and flake graphite. In other words, it is a thermionically assisted 2-photon emission process. We also performed curve fitting for the photoemission data of the copper photocathode, concluding that the temperatures under maximum laser irradiation for $J_0$, $J_1$ and $J_2$ have already exceeded the melting temperature of copper (1403 K)[28]. Under 1064nm-laser irradiation, $J_3$ is the most probable term to describe the photoemission mechanism of the copper photocathode under our experimental conditions. In other words, the emission mechanism of our copper photocathode at 1064 nm is a thermionically assisted 3-photon emission process. Table 3 enumerates the fitting parameters of the best $J_n$ terms for explaining the 1064-nm-laser-induced photoemission from the graphite and copper photocathodes. In our case, inducing photoemission from copper using an infrared laser requires laser heating near copper's melting temperature.

TABLE III. Fitting parameters for the photocurrents in Fig. 7.

|  | $a_n[(\frac{cm^2}{MW})^n]$ | $\eta[\frac{Kcm^2}{MW}]$ | Max. T [K] | Fitting Variance |
|---|---|---|---|---|
| $J_2$ (bulk) | 1.57 | 273 | 2400 | 1.42×10⁻¹ |
| $J_2$ (sheet) | 3.59×10² | 143 | 1473 | 4.56×10⁻² |
| $J_2$ (flake) | 2.46×10² | 352 | 1837 | 1.11×10⁻¹ |
| $J_3$ (copper) | 5.11×10⁻² | 76.0 | 925 | 1.10×10⁻³ |

Fig. 8 plots (a) the temperature and (b) quantum efficiency of the bulk-graphite, sheet-graphite, flake-graphite, and copper photocathodes versus laser intensity at 1064 nm. In Fig. 8(a), the cathode temperatures are all lower than the sublimation points or melting point of the cathode materials. However, the cathode temperature increases with the irradiating laser intensity, indicating a thermionically assisted photoemission process. In Fig. 8(b), the quantum efficiency of the flake-graphite photocathode exceeds all other materials under the same working laser intensity, although the photocurrent of the flake graphite saturates quickly at the highest value. At saturation, the measured maximum quantum efficiencies for flake, bulk and sheet graphite, as well as copper, are $5.1 \times 10^{-6}$, $1.0 \times 10^{-6}$, $8.1 \times 10^{-7}$, and $6.7 \times 10^{-9}$, respectively. The quantum efficiency of a graphite photocathode is 2-3 orders of magnitude higher than that of a copper photocathode irradiated by a 1064-nm pulse laser.

With the dominant fitting term $J_2$ (two-photon photoemission) for all graphite materials, the $a_2$ value for sheet graphite 358 $(\frac{cm^2}{MW})^2$ is relatively high when compared with 246 $(\frac{cm^2}{MW})^2$ for flake graphite and 1.57 $(\frac{cm^2}{MW})^2$ for bulk graphite. However, the maximum temperature 2400 K for the bulk-graphite cathode is higher than the 1837K for the flake one and the 1473K for the sheet one. This implies that, at saturation, thermionic emission is more pronounced for bulk graphite than flake graphite and sheet graphite irradiated by the infrared laser. Therefore, the superior emission ability of the flake-graphite photocathode is attributable to some field-emitted electrons from the protruding graphene flakes on the cathode surface. The effective surface area of the flake graphite is relatively large for trapping photons and emitting electrons, resulting in a high saturation current at a low irradiating laser intensity. The moderate temperature of the flake-graphite cathode in Fig. 8(a) contributes to an extended photocathode lifespan.

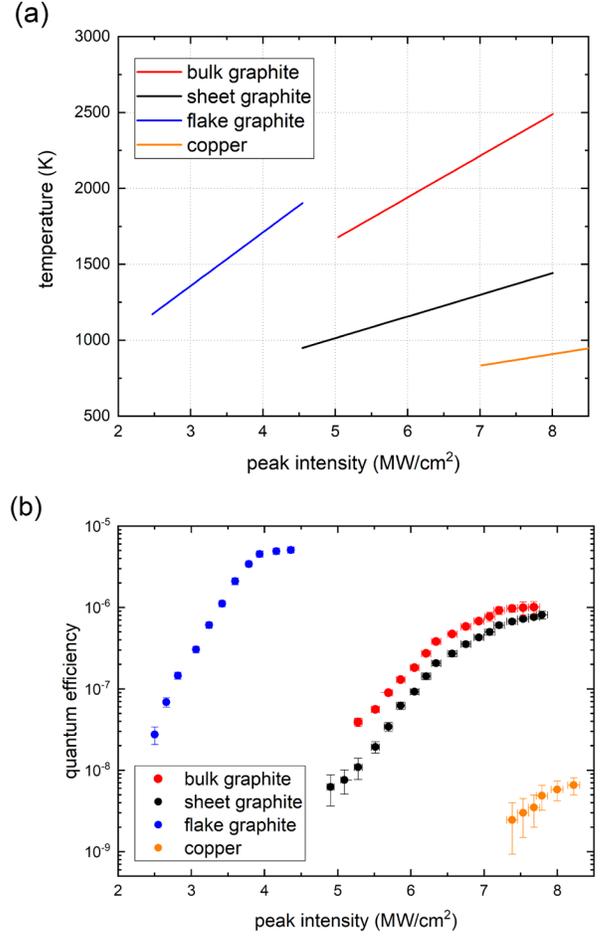

FIG. 8. (a) Cathode temperature and (b) quantum efficiency vs. laser intensity at 1064 nm for the bulk-graphite (red), sheet-graphite (black), flake-graphite (blue), and copper (orange) photocathodes. The flake-graphite photocathode shows superior quantum efficiency without laser damage to the cathode.

A high emission current or a high quantum efficiency does not necessarily guarantee a high brightness for a photoemitter. The temperature-dependent emittance of thermionically emitted electrons affects the brightness of a beam. Fig. 9 plots (a) the spot-normalized emittance, $\epsilon_N/\sigma_x$, and (b) the corresponding calculated beam brightness versus laser intensity for the 1064-nm laser irradiated bulk-graphite, sheet-graphite, flake-graphite, and copper photocathodes. Although the thermal emittance of the graphite photocathodes is higher than that of the copper photocathode, a graphite photoemitter is brighter due to a much higher photocurrent at the same laser irradiating intensity at 1064 nm. With the emittance values of 0.64, 0.50, and 0.56 mm·mrad/mm for the bulk- graphite, sheet- graphite, and flake-graphite emitters at the saturation current densities, the maximum brightness values of the emitters are 1.0, 1.6, and 3.9 A/(mm·mrad)² under the laser-irradiation intensities of 7.7, 8.2, and 4.4 MW/cm², respectively. In comparison, with an emittance value of 0.46 mm·mrad/mm at a laser-irradiation intensity of 8.2 MW/cm² for the copper photocathode, the maximum brightness of the copper photoemitter is 0.02A/(mm·mrad)², which is about 200 times lower than that of the flake-graphite photocathode. Furthermore, the brightness value of 3.9 A/(mm·mrad)² for the flake-graphite photoemitter is 11 times greater than the nominal value 0.35A/(mm·mrad)² typically quoted for the brightness of a high-energy photoinjector adopting a UV-laser irradiated copper photocathode[29]. The brightness of the graphite photocathodes is further comparable to the typical brightness of 2A/(mm·mrad)² for a LaB$_6$ field emitter[30] for low-energy electron microscopes.

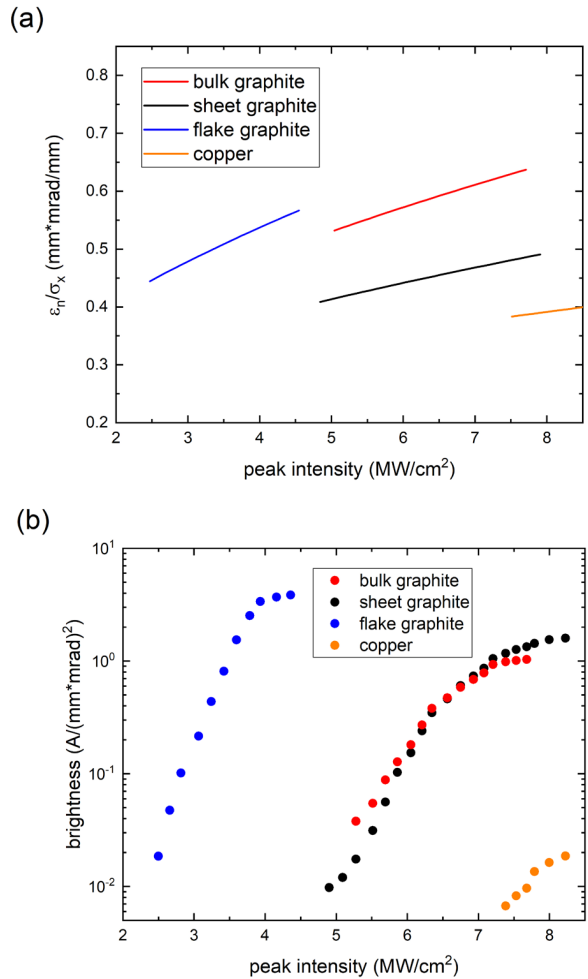

**FIG. 9.** (a) Spot-normalized emittance, and (b) brightness vs. laser intensity at 1064 nm for bulk-graphite (red), sheet-graphite (black), flake-graphite (blue), and copper (orange) photocathodes. Despite higher thermal emittance for the graphite photocathode, for space-charge limited emission, the flake-graphite photoemitter is 200 times brighter than a copper photoemitter excited by an infrared laser at 1064 nm.

## V. CONCLUSION

An infrared-laser-excited photoemitter operating in a moderate vacuum with high brightness, long lifetime, and low cost is highly desirable for many applications using an electron beam. In this paper, we report graphite as a promising photocathode material for a desirable photoemitter, even though the work function of graphite between 3.6-3.9 eV is significantly larger than the irradiating photon energy of our infrared lasers at 1064 nm. We experimentally investigated photoemission from 3 different types of graphite, including bulk graphite, sheet graphite, and flake graphite, in a 10$^{-6}$-torr vacuum chamber. Bulk graphite is fabricated by hot-pressing graphite powders to form a solid with a low level of crystallinity. Sheet graphite contains a thin stack of aligned graphene layers. The flake graphite is a novel form of graphite that we fabricated on a conducting surface with a dense population of nanographene fins protruding upward. Our study of the graphite photocathodes compares favorably with a copper photocathode in the same experimental setup.

We unfolded the photoemission mechanisms of the three graphite cathodes by fitting the experimentally measured photocurrents with the Fowler-DuBridge-Bechtel theory, showing a close alignment between experiment and theory for an incident photon energy lower than the work function of graphite. With 1064-nm laser irradiation, thermionically assisted 2-photon absorption is the dominant emission mechanism of electrons from the bulk, sheet, and flake graphite. Under the same experimental setup, a copper photocathode emits electrons via a thermionically assisted 3-photon emission process with relatively poor efficiency.

Graphite strongly absorbs visible light. We also tested the bulk-graphite photocathode at the second harmonic wavelength of our 1064-nm laser. With 532-nm laser irradiation, electron generation from bulk graphite is dominated by a thermionically assisted 1-photon emission process. Although the threshold of photoemission from graphite at 532 nm is relatively low, the operation of the cathode is limited to laser-induced damage at an intensity of ~5 MW/cm². On the other hand, when we irradiated the bulk-graphite photocathode with an infrared-laser intensity of 7.5 MW/cm² at 1064 nm, we detected a space-charge-limited photocurrent of 6.5 A/cm² without observing any damage to the cathode surface.

The surface texture of flake graphite is particularly efficient for trapping photons and emitting electrons. With an irradiating laser at 1064 nm, the space-charge-limited current density of the flake-graphite photocathode is approximately 200 times higher than that of the copper photocathode. The corresponding quantum efficiency of the flake-graphite photocathode is 5×10$^{-6}$, which is 770 times that of the copper photocathode tested in the same experiment and comparable to that of a typical copper photocathode driven by a UV laser at ~250 nm.

Electron emittance is an important parameter for calculating the brightness of an electron emitter. By applying Fowler's theory and assuming a Fermi gas model, we present a theory for calculating the emittance of thermionically assisted photoemission. Combining the measured photocurrent and calculated emittance allows us to infer the brightness of an electron emitter, revealing that the brightness of our flake-graphite photocathode is 200 times that of the copper photocathode at a laser wavelength of 1064 nm and 2 times that of a typical LaB$_6$ field emitter. Our study also shows that an infrared laser-irradiated flake-graphite photocathode is 10 times brighter than a typical UV-laser-irradiated copper photocathode. The findings in this study highlight the tremendous potential of a low-cost graphite-based photocathode that operates with an infrared laser in a moderate vacuum. In particular, among the tested graphite photocathodes, the flake-graphite photocathode exhibited superior photoemission, quantum efficiency, and brightness.

## AUTHOR DECLARATIONS

### Conflict of Interest

The authors declare no conflict of interest.

## DATA AVAILABILITY

The data that support the findings of this study are available from the corresponding author upon reasonable request.

## ACKNOWLEDGMENTS

This work is supported by National Science and Technology Council of Taiwan under Contract NSTC 112-2112-M-007 -028 -MY3. The authors would like to thank Ms. Ching-Wen Tsai for her assistance with UPS measurements.

## REFERENCES

[1] A. H. Zewail, "Four-dimensional electron microscopy," Science, 328, 187–193 (2010)

[2] R. Alley et al., "The design for the LCLS RF photoinjector," Nucl. Instrum. Methods Phys. Res., Sect. A 429, 324–331 (1999).

[3] S. P. Weathersby et al., "Mega-electron-volt ultrafast electron diffraction at SLAC National Accelerator Laboratory," Rev. Sci. Instrum. 86, (2015).

[4] P. A. Anderson, "The work function of copper," Phys. Rev. 76, 388 (1949).

[5] L. B. Jones et al., "Mean transverse energy, surface chemical and physical characterization of CERN-made Cs-Te photocathodes," Phys. Rev. Accel. Beams 27, 023402 (2024).

[6] J. Biswas et al., "High quantum efficiency GaAs photocathodes activated with Cs, O$_2$, and Te," AIP Adv. 11, (2021).

[7] X. Wang et al., "NEA GaAs photocathode for electron source: From growth, cleaning, activation to performance," Mater. Today Phys. 101680, (2025).

[8] T. Hirano, K. E. Urbanek, A. C. Ceballos, D. S. Black, Y. Miao, R. J. England, R. L. Byer, and K. J. Leedle, "A compact electron source for the dielectric laser accelerator," Appl. Phys. Lett. 116, 161106 (2020).

[9] A. I. Savvatimskiy, "Measurements of the melting point of graphite and the properties of liquid carbon (a review for 1963–2003)," Carbon 43, 1115–1142 (2005).

[10] M. I. Yalandin et al., "Study of explosive-emissive graphite cathodes in the pulsed-periodic operating regime," Proc. 21st Int. Symp. Discharges Electr. Insul. Vac. 2, IEEE (2004).

[11] D. Gugel et al., "Two-photon photoemission from image-potential states of epitaxial graphene," 2D Mater. 2, 045001 (2015).

[12] R. H. Fowler, "The analysis of photoelectric sensitivity curves for clean metals at various temperatures," Phys. Rev. 38, 45 (1931).

[13] L. A. DuBridge, "A further experimental test of Fowler's theory of photoelectric emission," Phys. Rev. 39, 108 (1932).

[14] L. A. DuBridge, "Theory of the energy distribution of photoelectrons," Phys. Rev. 43, 727 (1933).

[15] J. H. Bechtel, W. L. Smith, and N. Bloembergen, "Four-photon photoemission from tungsten," Opt. Commun. 13, 56–59 (1975).

[16] J. H. Bechtel, W. L. Smith, and N. Bloembergen, "Two-photon photoemission from metals induced by picosecond laser pulses," Phys. Rev. B 15, 4557 (1977).

[17] B. N. Chichkov et al., "Femtosecond, picosecond and nanosecond laser ablation of solids," Appl. Phys. A 63, 109–115 (1996).

[18] P. Debye, "Zur theorie der spezifischen wärmen," Ann. Phys. 344, 789–839 (1912).

[19] C. Kittel, Introduction to Solid State Physics, 8th Ed. (John Wiley & Sons, Hoboken, NJ, 2004).

[20] A. Tari, The Specific Heat of Matter at Low Temperatures (World Scientific, Singapore, 2003).

[21] G. R. Stewart, "Measurement of low-temperature specific heat," Rev. Sci. Instrum. 54, 1–11 (1983).

[22] C. Kittel and H. Kroemer, "Thermal physics," Am. J. Phys. 39, 126–127 (1971).

[23] R. K. Pathria, Statistical Mechanics: International Series of Monographs in Natural Philosophy, Vol. 45 (Elsevier, Amsterdam, 2017).

[24] D. H. Dowell and J. F. Schmerge, "Quantum efficiency and thermal emittance of metal photocathodes," Phys. Rev. ST Accel. Beams 12, 074201 (2009).

[25] P.-H. Wu and Y.-C. Huang, " Theory of Mean Transvers Energy including both Thermionic, Multiphoton, and Field emissions," to appear in Phys. Rev. Accel. Beams (2025).

[26] A. B. Djurišić and E. H. Li, "Optical properties of graphite," J. Appl. Phys. 85, 7404–7410 (1999).

[27] Y. Kawamura, K. Toyoda, and M. Kawai, "Generation of relativistic photoelectrons induced by excimer laser irradiation," Appl. Phys. Lett. 45, 307–309 (1984).

[28] W. J. Kroll, "Melting and evaporating metals in a vacuum," Trans. Electrochem. Soc. 87, 571 (1945).

[29] F. Stephan and M. Krasilnikov, "High brightness photo injectors for brilliant light sources," in Synchrotron Light Sources and Free-Electron Lasers: Accelerator Physics, Instrumentation and Science Applications, E. Jaeschke, S. Khan, J. R. Schneider, and J. Hastings, Eds. (Springer International Publishing, Cham, 2020), pp. 603–646.

[30] L. Reimer, "Elements of a transmission electron microscope," in Transmission Electron Microscopy: Physics of

Image Formation and Microanalysis (Springer, 1989), pp. 86–135.